\documentclass{article}

\usepackage[margin=1.25in]{geometry}
\usepackage[utf8]{inputenc}
\usepackage[multiple]{footmisc}
\usepackage{amsmath}
\usepackage{amssymb}
\usepackage{amsthm}
\usepackage{bbm}
\usepackage{breakcites}
\usepackage[T1]{fontenc}

\usepackage{multicol}
\usepackage{booktabs}
\usepackage{graphicx}
\usepackage{float}
\usepackage{bm}
\usepackage{multirow}
\usepackage{xargs}
\usepackage{listings}
\usepackage{xcolor}
\usepackage{tabularx}
\usepackage{algorithm}
\usepackage{algcompatible}

\usepackage{bbm}
\usepackage{comment}
\usepackage{natbib}
\usepackage[colorlinks=true, allcolors=blue]{hyperref}

\definecolor{codegreen}{rgb}{0,0.6,0}
\definecolor{codegray}{rgb}{0.5,0.5,0.5}
\definecolor{codepurple}{rgb}{0.58,0,0.82}
\definecolor{backcolour}{rgb}{0.95,0.95,0.92}
\lstdefinestyle{mystyle}{
    backgroundcolor=\color{backcolour},   
    commentstyle=\color{codegreen},
    keywordstyle=\color{magenta},
    numberstyle=\tiny\color{codegray},
    stringstyle=\color{codepurple},
    basicstyle=\ttfamily\footnotesize,
    breakatwhitespace=false,         
    breaklines=true,                 
    captionpos=b,                    
    keepspaces=true,                 
    showspaces=false,                
    showstringspaces=false,
    showtabs=false,                  
    tabsize=2
}
\restylefloat{table}

\renewcommand{\thesubsection}{\Alph{subsection}}
\title{Enhancing Longitudinal Clinical Trial Efficiency with Digital Twins and Prognostic Covariate-Adjusted Mixed Models for Repeated Measures (PROCOVA-MMRM)}
\author{Jessica L. Ross$^1$, Arman Sabbaghi$^1$, Run Zhuang$^1$, Daniele Bertolini$^1$,\\
for the ADCS\footnote{Data used in preparation of this manuscript/publication/article were obtained from the University of California, San Diego Alzheimer’s Disease Cooperative Study (https://www.adcs.org/)},
the Alzheimer’s Disease Neuroimaging Initiative\footnote{Data used in preparation of this article were obtained from the Alzheimer’s Disease Neuroimaging Initiative (ADNI) database (adni.loni.usc.edu). As such, the investigators within the ADNI contributed to the design and implementation of ADNI and/or provided data but did not participate in analysis or writing of this report. A complete listing of ADNI investigators can be found at: https://adni.loni.usc.edu/wp-content/uploads/how\_to\_apply/ADNI\_Acknowledgement\_List.pdf},
the CPAD \\Database\footnote{Data used in the preparation of this article were obtained from the Critical Path for Alzheimer’s Disease
(CPAD) Database. In 2008, Critical Path Institute, in collaboration with the Engelberg Center for Health Care Reform at the Brookings Institution, formed the Coalition Against Major Diseases (rebranded to Critical Path for Alzheimer’s Disease (CPAD) consortium in 2018). The consortium brings together patient groups, biopharmaceutical companies, and scientists from academia, the U.S. Food and Drug Administration (FDA), the European Medicines Agency (EMA), the National Institute of Neurological Disorders and Stroke (NINDS), and the National Institute on Aging (NIA). The Critical Path for Alzheimer’s Disease (CPAD) consortium includes over 200 scientists from member and non-member organizations. The data available in the CPAD database has been volunteered by CPAD member companies and non-member organizations.}, the European Prevention of Alzheimer’s Disease (EPAD) Consortium\footnote{Data used in preparation of this article were obtained from the Longitudinal Cohort Study (LCS), delivered by the European Prevention of Alzheimer’s Disease (EPAD) Consortium. As such investigators within the EPAD LCS and EPAD Consortium contributed to the design and implementation of EPAD and/or provided data but did not participate in analysis or writing of this report. A complete list of EPAD Investigators can be found at: https://ep-ad.org/wp-content/uploads/2020/12/202010\_List-of-epadistas.pdf},\\
and the Pooled Resource Open-Access ALS Clinical Trials Consortium\footnote{Data used in the preparation of this article were obtained from the Pooled Resource Open-Access ALS Clinical Trials (PRO-ACT) Database. As such, the following organizations and individuals within the PRO-ACT Consortium contributed to the design and implementation of the PRO-ACT Database and/or provided data, but did not participate in the analysis of the data or the writing of this report: Neurological Clinical Research Institute, MGH; Northeast ALS Consortium; Novartis; Prize4Life Israel; Regeneron Pharmaceuticals, Inc.; Sanofi; Teva Pharmaceutical Industries, Ltd.}}
\date{
    $^1${Unlearn.AI, Inc., San Francisco, CA}\\
    \today
    }
\begin{document}

\maketitle

\begin{abstract}
Clinical trials are critical in advancing medical treatments but often suffer from immense time and financial burden. Advances in statistical methodologies and artificial intelligence (AI) present opportunities to address these inefficiencies. Here we introduce Prognostic Covariate-Adjusted Mixed Models for Repeated Measures (PROCOVA-MMRM) as an advantageous combination of prognostic covariate adjustment (PROCOVA) and Mixed Models for Repeated Measures (MMRM). PROCOVA-MMRM utilizes time-matched prognostic scores generated from AI models to enhance the precision of treatment effect estimators for longitudinal continuous outcomes, enabling reductions in sample size and enrollment times. We first provide a description of the background and implementation of PROCOVA-MMRM, followed by two case study reanalyses where we compare the performance of PROCOVA-MMRM versus the unadjusted MMRM. These reanalyses demonstrate significant improvements in statistical power and precision in clinical indications with unmet medical need, specifically Alzheimer’s Disease (AD) and Amyotrophic Lateral Sclerosis (ALS). We also explore the potential for sample size reduction with the prospective implementation of PROCOVA-MMRM, finding that the same or better results could have been achieved with fewer participants in these historical trials if the enhanced precision provided by PROCOVA-MMRM had been prospectively leveraged. We also confirm the robustness of the statistical properties of PROCOVA-MMRM in a variety of realistic simulation scenarios. Altogether, PROCOVA-MMRM represents a rigorous method of incorporating advances in the prediction of time-matched prognostic scores generated by AI into longitudinal analysis, potentially reducing both the cost and time required to bring new treatments to patients while adhering to regulatory standards.
\end{abstract} \hspace{10pt}

\section{Introduction}
\label{sec:introduction}

Clinical trials are notoriously inefficient, requiring immense time and financial investment to introduce potentially life-saving treatments to the market \citep{Fogel2018}. In addition, 80\% of trials fail to meet their initial enrollment targets \citep{BM2020}, which can result in insufficient power. Sponsors can improve the precision and power of their trial by including more participants, but the associated increases in trial time and expense make alternative strategies for improving trial efficiency more appealing.

One such strategy is to utilize observed outcome information across all timepoints in a study (longitudinally) to estimate the average treatment effect at the final timepoint for the full trial population. The limitations of common Analysis of (Co)Variance (ANOVA/ANCOVA)-based methods in dealing with routine challenges in randomized clinical trials (RCTs), such as missing data, have become apparent, and in recent years, more sophisticated methods like Mixed Models for Repeated Measures (MMRM) and Generalized Estimating Equations (GEE) have grown in favor. While all statistical methods require some assumptions when considering the pattern of missing data/participant dropout, the assumptions required for ANOVA-based methods are more rigid than those required by MMRM and GEE, which can handle missing data without explicit imputation \citep{Garcia2017, EMA_missingdata}. Data can be missing completely at random (MCAR), where there is no pattern to the missing data. More often, trials include data that are missing at random (MAR), which means that the missing data can be explained by observed variables in the study \citep{Mallinckrodt2008}, such as baseline covariates. For example, patients with worse functional measures at baseline might be more likely to drop out of a clinical trial. As their missingness would be partially explained by a variable measured in the study, this would be considered MAR. MMRM allows for strict type I error control and unbiased estimation of treatment effects for both MAR and MCAR data \citep{Mallinckrodt2001,Schuler2022}, whereas the common practice of ANOVA with last observation carried forward (LOCF) and GEE can only be used safely with MCAR data \citep{Mallinckrodt2001,Garcia2017}. MMRM is effectively an extension of the linear regression model, in which the model is specified
\begin{itemize}
    \item for all observed measurements on all participants,
    \item with different intercepts and treatment effects at distinct timepoints, and
    \item with a specified structure on the covariance matrix for the random error term in the model.
\end{itemize}
MMRM is suitable under current regulatory guidance \citep{EMA_missingdata} and is less sensitive to dropout bias and provides more power over standard analysis methods for continuous endpoints, contributing to its popularity for the analysis of phase 2 and 3 RCTs \citep{Kabashima2020,Richeldi2022,Jastreboff2022}. 

Another strategy for increasing power in RCTs is through the use of covariate adjustment. Regulatory guidance recommends that only a small number of covariates be used for adjustment in the statistical analysis of clinical trials \citep{FDA2023,EMA2015}; but, determining the small set of covariates that can yield the greatest reduction in the variance of the treatment effect estimator can be challenging in the case of the many baseline measurements assessed in RCTs. \cite{schuler_increasing_2021} has shown that the optimal adjustment covariate in the estimation of a treatment effect would be the participant's expected outcome under control, conditional on the covariates. While a participant’s expected control outcome cannot be known at study onset, the use of artificial intelligence (AI) can yield powerful predictions of expected control outcomes, or prognostic scores \citep{Hansen2008}, conditional on baseline covariates. AI models can handle complex, high-dimensional data and identify patterns that might not be apparent through conventional statistical models. As the availability of rich longitudinal participant data from historical clinical trials and observational studies has increased, so too has the predictive ability of AI models trained on this data. In fact, the European Medicines Agency (EMA) recently qualified a novel methodology, prognostic covariate adjustment (PROCOVA), which can enhance the efficiency of phase 2 and 3 RCTs without adding bias or risk of type I error in the case of continuous endpoint analysis. The power gain afforded by PROCOVA can also be anticipated during trial planning to prospectively reduce sample size over traditional analysis methods \citep{EMAPROCOVA}.

Here, we introduce PROCOVA-MMRM, an extension of PROCOVA intended for phase 2 and 3 RCTs with longitudinal continuous endpoints. Whereas PROCOVA is used for single timepoint analyses, PROCOVA-MMRM integrates prognostic covariate adjustment into the longitudinal analysis of continuous endpoints measured at multiple timepoints. Through the inclusion of time-matched prognostic scores, generated for each trial participant from only their baseline characteristics, as adjustment covariates, PROCOVA-MMRM can yield the following desirable results:
\begin{itemize}
    \item [] Improved precision of treatment effect inferences compared to traditional MMRM analyses for a fixed sample size that either have no adjustment or that adjust solely for baseline covariates, and
    \item[] Reductions in RCT enrollment time and control arm sample size without sacrificing trial power.
\end{itemize}

Improvements in this framework for longitudinal outcomes have mirrored advances in prognostic modeling. The Conditional Restricted Boltzmann Machines (CRBMs) \citep{Fisher2019,Bertolini2020} that were used to support PROCOVA predicted at a 3 month cadence, so their utility with longitudinal outcomes was limited to studies with the same visit structure. \cite{Lang2023} and \cite{Walsh2024} recently introduced the Neural Boltzmann Machine (NBM) architecture as an extension of CRBMs that allows for a more flexible representation of the underlying data. The NBM model can handle missing data and generate prognostic scores at arbitrary timepoints (rather than fixed timepoints), making it well-equipped to be used in trials that longitudinally collect continuous outcome measurements. Here, we demonstrate the efficiency gain achievable over traditional methods through the use of time-matched prognostic scores (i.e., predictions of control outcomes for an individual trial participant at planned observation timepoints) generated by NBMs for covariate adjustment in PROCOVA-MMRM. We begin with a description of the general methodology and step-by-step implementation of PROCOVA-MMRM, followed by case studies utilizing NBMs for Alzheimer’s disease (AD) and amyotrophic lateral sclerosis (ALS) to demonstrate the practical implementation of this method, either for power gain or prospective sample size reduction, in clinical indications with unmet medical needs. Finally, we perform a series of simulations to demonstrate the robustness of the procedure across a variety of scenarios not encountered in our case studies.

\section{The PROCOVA-MMRM Model}
\label{sec:procova_mmrm_background}

\subsection{Notations and Assumptions}
\label{sec:notations}

We consider an RCT with $N$ participants who are randomized to either the active treatment arm (denoted by $w_i=1$ for participant $i$) or the control arm (denoted by $w_i=0$). The vector of time-matched prognostic scores for each participant $i$  is denoted by $\mathbf{x}_i$, with entry $x_{i,t} \in \mathbb{R}$ for time $t$ determined from observations measured prior to treatment assignment. The continuous endpoint $y_{i,t} \in \mathbb{R}$ is measured at each time point $t = 1, \ldots, T_i$ for participant $i$, where $T_i \in \{ 1, \ldots, T \}$ is the final time point for which participant $i$ has an observed outcome. Note that $T_i$ is specified because of the potential for missing data, thus the last observed timepoint for participant $i$ could differ from the last observed timepoint for another participant. We let $\mathbf{y}_i=(y_{i,1}, \ldots, y_{i,T_i})'$ denote the vector of observed outcomes for participant $i$. The trial dataset is the set of tuples $\{w_i,\mathbf{x}_i,\mathbf{y}_i:i=1,\ldots,N\}$. 

PROCOVA-MMRM is a form of the MMRM-II working model (with interactions) described by \cite{WangDu2023}. This model is
\begin{equation}
\label{eq:mmrm_II}
\left [ \mathbf{y}_i \mid w_i, \mathbf{x}_i \right ] \sim \mathrm{Normal}_{T_i} \left ( \begin{pmatrix} \beta_{0,1} \\ \vdots \\ \beta_{0,T_i} \end{pmatrix} + w_i \begin{pmatrix} \beta_{w,1} \\ \vdots \\ \beta_{w,T_i} \end{pmatrix} + \mathbf{B}_{x}\mathbf{x}_i, \mathbf{R}_i \right )
\end{equation}
where $\mathbf{B}_x = \begin{pmatrix} \boldsymbol{\beta}_{x,1}' \\ \boldsymbol{\beta}_{x,2}' \\ \ldots \\ \boldsymbol{\beta}_{x,T_i}' \end{pmatrix}$ is the $T_i \times T_i$ matrix of unknown model coefficients in which each row consists of a distinct set of parameters indexed by time for the $T_i$-dimensional vector of time-matched prognostic scores and $\mathbf{R}_i$ is the $T_i \times T_i$ positive definite covariance matrix for the random error term of participant $i$. The predictor vectors in this model are effectively defined in terms of the indicators for the timepoints $\mathbb{I}_t$, the products of the timepoint indicators with the treatment indicator $w_i\mathbb{I}_t$, and the products of the timepoint indicators with the prognostic scores $x_{i,t}\mathbb{I}_t$. Here, the indicator vector $\mathbb{I}_t$ is a $T \times 1$ vector that has a $1$ in entry $t$, and zeros for all other entries.

It is important to recognize that, in this model, the time-matched prognostic scores are not a function of treatment nor of any intermediate outcomes, and are purely a transformation of the baseline covariates via a prognostic model that can be calculated and observed at baseline. Consequently, the prognostic scores are not time-varying covariates \citep{Diggle2013}, but are equivalent to a vector of baseline covariates.

PROCOVA-MMRM proceeds by fitting an MMRM model with the adjustment for time-matched prognostic scores $x_{i,t}$ (including interactions with the time indicator) to the data using Restricted Maximum Likelihood (REML) with an unstructured covariance matrix. Robust standard errors are utilized for the treatment effect estimator. Once the model has been fitted, the point estimate, p-value, and confidence interval for $\beta_{w,T}$ (which is the estimand of interest) are calculated.

\subsection{Sample Size Estimation}
\label{sec:sample_size_estimation}

PROCOVA-MMRM yields unbiased treatment effect estimators with desired interval coverage rates and controlled type I error rates in the large sample setting. Closed-form prospective sample size reduction formulae for the MMRM-II working model do not exist. Using simulation studies, efficiency gains (i.e., power boost and improved precision for treatment effect estimation) using PROCOVA-MMRM can be quantified with respect to the traditional MMRM analyses that either do not adjust for any covariates or that adjust solely for baseline covariates. This comparison using specific trial characteristics enables the fullest benefit of PROCOVA-MMRM for sample size reduction to be quantified.

As an alternative, one may use the PROCOVA formula for sample size reduction \citep{EMAPROCOVA}. PROCOVA-MMRM will always achieve a lower variance of the treatment effect estimator over complete-case PROCOVA due to the additional longitudinal information incorporated into the model (Appendix \ref{sec:appA}). Thus, the PROCOVA sample size estimation formula can also be utilized as a conservative estimate for sample size in PROCOVA-MMRM. This implementation of PROCOVA-MMRM is described in the next section.

\section{Implementation for Prospective Trial Planning}
\label{sec:procova_mmrm_implementation}

\subsection{Step 1: Validate the Prognostic Score for Use in the Target Trial}
\label{sec:step_1}

The purpose of Step 1 is to validate the prognostic score generated by a prognostic model for use in a particular planned trial, which we will call the Target Trial. In the case that one plans to use the PROCOVA sample size estimation formula as a conservative estimate, the validation will require estimating the Pearson correlation coefficient $R$ between the end-of-study prognostic score and the actual outcomes obtained from a separate dataset that was not used to train the prognostic model, which contains data from subjects whose baseline characteristics are similar to those in the Target Trial.

\subsection{Step 2: Estimate Sample Size and Plan the Target Trial Taking the Prognostic Score into Account}
\label{sec:step_2}

\subsubsection{Step 2a: Estimate Sample Size}
\label{sec:step_2_a}

Gather the standard inputs needed to compute a sample size for a given power (i.e., the target effect size, the standard deviation of the outcome, the proportion of subjects to be randomized to the intervention, the expected dropout rate, and the alpha level), and define the inflation and deflation factors used to avoid undue optimism in the PROCOVA sample size calculation, i.e., $\lambda$ (the deflation factor for the correlation coefficient $R$) and $\gamma$ (the inflation factor for the standard deviation). Use the rules outlined in the PROCOVA Handbook for the Target Trial Statistician to select $\lambda$ and $\gamma$ \citep{EMAPROCOVA}. With all of the parameters now defined, compute the power over a range of sample sizes to obtain a power curve, and choose the minimum sample size to achieve the desired power for the Target Trial. In general, the PROCOVA sample size formula does not require a 1:1 randomization and allows for the $\lambda$ and $\gamma$ values to differ between the control and treatment groups. Here, we provide an example of a simplified formula where a common $\gamma$ and a common $\lambda$ are used for two equally sized treatment groups. We also assume that the outcomes have a common variance $\sigma^2$; that $n$  represents the trial sample size without dropouts; and that $d$ represents the proportion of subjects who drop out.
\begin{itemize}
    \item Estimate the standard error $\nu$ of the treatment effect via $\sqrt{\left \{ n(1-d) \right \}^{-1} (2\gamma\sigma)^2 \left \{ 1 - \left ( \lambda R \right )^2 \right \}}$.
    \item Compute the power as $\Phi(\Phi^{-1}(\alpha/2)+\beta/\nu)+\Phi(\Phi^{-1}(\alpha/2)-\beta/\nu)$, where $\Phi \left ( \cdot \right )$ is the cumulative distribution function of the standard Normal distribution, $\alpha$ is the type I error rate, and $\beta$ is the target treatment effect.
    \item For a range of $n$ (e.g., 100-1000) compute the power and plot the corresponding curve.
    \end{itemize}
If the Target Trial utilizes co-primary endpoints, repeat the bulleted steps above for each endpoint. Select the sample size that would be sufficient to simultaneously address the null hypothesis for both endpoints. Note that repeating the process for multiple endpoints necessitates having a prognostic score for each endpoint, and that, in turn, necessitates either a machine learning model that predicts multivariate results or multiple individual models for each endpoint of interest.

\subsubsection{Step 2b: Plan the Target Trial and Pre-Specify its Characteristics}
\label{sec:step_2_b}

To apply PROCOVA-MMRM, a validated prognostic model used to generate the time-matched prognostic scores for the Target Trial must be locked and prespecified in the Target Trial protocol and/or statistical analysis plan (SAP) as appropriate, prior to implementation, to prevent data-driven model selection. This documentation should also include model weights and any hyperparameters associated with the sampling necessary to compute the prognostic scores.
\begin{itemize}
    \item Consider if individual baseline covariates will be included (along with the time-matched prognostic scores) in the primary analysis and/or sensitivity analysis. When PROCOVA-MMRM is applied to trials utilizing stratified randomization, the strata should be included as covariates in the primary analysis (note that the prognostic scores are not designed to be used for stratification).
    \item Choose the strategy for handling dropout by defining the appropriate estimand. For example, the estimand could be treatment policy or hypothetical depending on how missing data are handled, even though both methods can use PROCOVA-MMRM.
    \item Select and pre-specify sensitivity analyses that probe the assumptions of PROCOVA-MMRM, as you would with MMRM alone. To evaluate the ability of PROCOVA-MMRM to address missing data, methods such as MICE \citep{white2011} or tipping point analyses \citep{Liublinska2014} can be applied. Additional sensitivity analyses for PROCOVA-MMRM may include principal stratification \citep{EMA_Estimands}, unadjusted MMRM, and/or MMRM with baseline covariates, dependent on Trial Statistician preference.
    \item Although PROCOVA-MMRM uses an unstructured covariance matrix, select and pre-specify at least two backup options for the structure of the covariance matrix to account for the possibility that the unstructured covariance matrix fails to converge. We recommend using a Toeplitz structure with an additional backup of a compound symmetry covariance structure in the case that both the unstructured and Toeplitz matrices fail to converge.
    \item Document all analysis choices described in Step 2b in the Target Trial protocol and/or SAP, as appropriate, prior to implementation.
\end{itemize}

\subsection{Step 3: Analyze Target Trial Results Using MMRM That Adjusts for the Time-Matched Prognostic Scores.}
\label{sec:step_3}

\begin{itemize}
    \item [] \textbf{Step 3a:} When clean baseline data on all participants randomized into the Target Trial are available, provide them to the model developers for the generation of trial-specific prognostic scores. The model developers must remain blinded to the randomization code until after the final trial-specific prognostic score is delivered to the Target Trial Statistician and applied in treatment effect estimation as described below.
    \item[] \textbf{Step 3b:} Estimate the treatment effect with PROCOVA-MMRM using MMRM that adjusts for the time-matched prognostic scores. If specified in the protocol and SAP, adjust for additional covariates (including stratification factors in the case of trials with stratified randomization). In all cases, estimate the variance for the treatment effect using a heteroskedasticity-consistent (HC) covariance matrix. In SAS, this is the empirical option in PROC MIXED which computes the Huber-White covariance of the fixed parameters \citep{Huber1967,White1980}. To achieve similar results, the implementation in R uses the \texttt{mmrm} package with the Kenward-Roger procedure to adjust degrees of freedom. The treatment effect estimator corresponds to the estimator of $\beta_{w,T}$ from the fit of the PROCOVA-MMRM model to the entire data. The treatment effect estimate, as well as the standard error associated with the estimate, are obtained in SAS via the LSMEANS statement.
    
    For this primary analysis, test the null hypothesis by computing a two-sided p-value based on a t-distribution using the covariate-adjusted treatment effect and the HC standard error.
\end{itemize}
    
\section{Case Studies}
\label{sec:case_studies}

Recruitment and enrollment are known challenges for Phase 2 and 3 RCTs in both AD and ALS. Furthermore, sufficient historical data (from RCTs and observational studies) exist that can be used to train prognostic models for each of these conditions. Thus, to illustrate the broad utility of PROCOVA-MMRM across therapeutic indications where highly predictive prognostic models are available, we reanalyze two previously completed phase 2/3 clinical trials in individuals with AD or ALS.

\subsection{General Methods}
\label{sec:case_studies_general_methods}

The AI models for AD and ALS that were used to generate time-matched prognostic scores for trial participants in these case studies utilized a newer model architecture that improved upon the previous CRBMs \citep{Fisher2019,Bertolini2020}, which were limited by their underlying representational power and restriction of prediction times. To overcome these limitations, we generalized CRBMs to Neural Boltzmann Machines (NBM), which promote parameters of a CRBM to full neural networks and allow for greater flexibility. A comprehensive discussion of the model architecture is outside the scope of this manuscript and we refer to \cite{Lang2023} for details. In brief, the main objective of the model is to predict a multivariate distribution of outcomes over time, when prompted with the measured baseline characteristics of a given trial participant. Two separate neural networks, which take as input baseline characteristics, predict the mode and the per-participant precision of the multivariate distribution at arbitrary followup times. Trajectories for a given trial participant (i.e., the samples forming their digital twin distribution) are then generated autoregressively, where predictions at a future time $t_{\mathrm{future}}$ are produced from baseline and from predictions at the previous time $t_{\mathrm{current}}$. A third neural network, a linear auto-encoder, imputes missing baseline values and it is co-trained with the NBM. This new architecture enables greater flexibility in missing data handling and prediction cadence, making it particularly powerful for generating time-matched prognostic scores.

Rather than a single fixed 50\% / 20\% / 30\% split into training, validation, and test datasets as was performed with previous versions of our model for AD progression \citep{Fisher2019,Bertolini2020}, the filtered training data was divided into five folds of similar size according to the following strategy: clinical trials were assigned to individual folds while observational studies were split across folds and stratified by baseline severity and other basic demographic variables. Assigning clinical trials to individual folds makes cross-validation more robust and improves estimation of how performance will generalize to future trials. On the other hand, for observational studies, which are typically larger, stratification ensures that the severity distribution in each fold roughly matches the distribution over the full population. For each run of a potential model, four of the folds were used for training and one was held out for evaluation of validation performance. This process was repeated until each fold had been held out for evaluation once, and the mean and standard deviation of these five evaluations of independent model training were used to represent the spread of possible performance of the prognostic model. In order to form a more robust prognostic model for real-world use cases, we use these five independently trained prognostic models to form an ensemble sampler, with each model contributing one fifth of the requested samples that form the digital twin distribution for each individual trial participant. This cross-validation procedure was performed for both the AD and ALS prognostic models.

The first reanalysis was performed using a phase 3 study investigating the effect of docosahexaenoic acid (DHA) on cognitive function in patients with AD \citep{Quinn2010}, to be referred to hereafter as the AD Target Trial. The second reanalysis was conducted on a phase 2/3 study \citep{Cudkowicz2014} investigating the effect of ceftriaxone on function and survival in patients with ALS, to be referred to hereafter as the ALS Target Trial (NCT00349622). Importantly, neither of these Target Trial data sets were used in model training. For each of the trials, we present the results of two experiments:
\begin{enumerate}
    \item [] \textbf{Experiment 1.} Pre-specified primary analysis of phase 2 and 3 trials, to deliver higher power/confidence in the results compared to unadjusted analyses.
    \item [] \textbf{Experiment 2.} Prospective design/sample size estimation for phase 2 and 3 trials, to attain the desired level of power/confidence with a smaller sample size compared to unadjusted designs. This follows the implementation procedure using PROCOVA sample size estimation as described in Section \ref{sec:procova_mmrm_implementation}.
\end{enumerate}

We show that PROCOVA-MMRM yields smaller estimated variances of the treatment effect estimators in each of the re-analyzed studies, regardless of whether a conservative reduction in sample size (via PROCOVA sample size estimation) is prospectively applied. The MMRM and PROCOVA-MMRM models were fit using PROC MIXED in SAS Version 9.4 for LIN64 (SAS Institute, Cary, NC). The robust heteroskedasticity-consistent variance estimator was used to perform statistical tests using $\alpha = 0.05$. Validity of the assumptions for the analyses were assessed via standard diagnostics.

\subsection{Case Study 1 - Alzheimer's Disease}
\label{sec:case_study_AD}

Data from the AD Target Trial were obtained from the University of California, San Diego through the Alzheimer's Disease Cooperative Study (ADCS), a consortium of academic medical centers and private AD clinics funded by the National Institute on Aging to conduct clinical trials on AD (NCT00440050). In this trial, 238 participants were randomized to the active treatment arm, and 164 participants were randomized to placebo (3:2 randomization). The AD Target Trial measured multiple baseline characteristics including demographics (e.g., sex, age, weight, geographic region), lab tests (e.g., blood pressure, ApoE4 status) \citep{Coon2007,safieh2019}, and component scores of cognitive tests. Our reanalysis was performed on both co-primary endpoints in this trial: the 18-month change in the Alzheimer’s Disease Assessment Scale–Cognitive Subscale 11 (ADAS-Cog 11) score \citep{Rosen1984} and the 18-month change in the Clinical Dementia Rating Scale Sum of Boxes (CDR-SB) score \citep{Morris1993}.

Along with the change in model architecture from CRBM to NBM, this version of the prognostic model included additional variables and was trained on a significantly larger historical dataset. A dataset of 25,212 participants, covering 76 background and longitudinal variables, was used to train the prognostic model to predict control outcomes of a trial participant given their baseline characteristics. These historical data were obtained from the National Alzheimer's Coordinating Center (NACC) \citep{Besser2018}, the Knight Alzheimer’s Disease Research Center \citep{koenig2020}, and three additional data sources. Data used in the preparation of this article were obtained from the Alzheimer’s Disease Neuroimaging Initiative (ADNI) database (adni.loni.usc.edu). The ADNI was launched in 2003 as a public-private partnership, led by Principal Investigator Michael W. Weiner, MD. The primary goal of ADNI has been to test whether serial magnetic resonance imaging (MRI), positron emission tomography (PET), other biological markers, and clinical and neuropsychological assessment can be combined to measure the progression of mild cognitive impairment (MCI) and early AD \citep{Petersen_2010}. Data used in the preparation of this article were obtained from the Critical Path for Alzheimer’s Disease (CPAD) Database. As such, the investigators within CPAD contributed to the design and implementation of the CPAD database and/or provided data, but did not participate in the analysis of the data or the writing of this report \citep{Romero2009,Neville2015}. Data used in preparation of this article were obtained from the EPAD Longitudinal Cohort Study (LCS) dataset V.IMI, doi:10.34688/epadlcs\_v.imi\_20.10.30. The EPAD LCS was launched in 2015 as a public-private partnership, led by Chief Investigator Professor Craig Ritchie MB BS. The primary research goal of the EPAD LCS (NCT02804789) is to provide a well-phenotyped probability-spectrum population for developing and continuously improving disease models for Alzheimer’s disease in individuals without dementia \citep{solomon2018}. Importantly, the training dataset did not include data from the AD Target Trial.

\subsubsection{AD Experiment 1}
\label{sec:case_study_AD_1}

After fitting the prognostic model and obtaining time-matched prognostic scores for each study visit, we analyzed the results from the AD Target Trial using four approaches: MMRM; MMRM with covariate adjustment for baseline (BL) ADAS-Cog-11 or CDR-SB score; PROCOVA-MMRM; and PROCOVA-MMRM with adjustment for BL ADAS-Cog-11 or CDR-SB score as an additional covariate. These reanalyses used the same number of participants and randomization ratio as in the original analysis reported by \cite{Quinn2010}. The point estimates and variance of the treatment effect for ADAS-Cog11 (Table \ref{tab:AD-Experiment1-ADAS}) and CDR-SB (Table \ref{tab:AD-Experiment1-CDR}) at 18 months were obtained in each of the four analysis conditions. All variance data presented are heteroskedasticity consistent and were computed using the Huber-White estimator. 

\begin{table}[H]
    \centering
    \begin{tabular} {|l|c|c|}
    \hline
         \textbf{Analysis Method} & \textbf{Treatment Effect Estimate} & \textbf{Estimated Variance} \\
          \hline
         Unadjusted MMRM & -0.035 & 1.024 \\  
         MMRM + BL ADAS-Cog11 & -0.028 & 0.955 \\  
         PROCOVA-MMRM & 0.158 & 0.907 \\  
         PROCOVA-MMRM + BL ADAS-Cog11 & 0.117 & 0.857 \\   
         \hline
    \end{tabular}
    \caption{Estimated treatment effects and variances for ADAS-Cog11 at month 18 in the reanalysis of the AD Target Trial.}
    \label{tab:AD-Experiment1-ADAS}
\end{table}

\begin{table}[H]
    \centering
    \begin{tabular} {|l|c|c|}
    \hline
         \textbf{Analysis Method} & \textbf{Treatment Effect Estimate} & \textbf{Estimated Variance} \\
          \hline
         Unadjusted MMRM & -0.092 & 0.113 \\  
         MMRM + BL CDR-SB & -0.022 & 0.108 \\  
         PROCOVA-MMRM & -0.116 & 0.099 \\  
         PROCOVA-MMRM + BL CDR-SB & -0.038 & 0.093 \\   
         \hline
    \end{tabular}
    \caption{Estimated treatment effects and variances for CDR-SB at month 18 in the reanalysis of the AD Target Trial.}
    \label{tab:AD-Experiment1-CDR}
\end{table}

We also evaluated how well the outcomes predicted by our prognostic model correlated with the subsequent study observations. For ADAS-Cog11, the correlation between the time-matched prognostic scores generated from baseline data and the study visit observations was 0.36, 0.34, and 0.38 at 6, 12, and 18 months, respectively. For CDR-SB, the correlation between the time-matched prognostic scores predicted from baseline data only and the actual outcomes was 0.34, 0.37, and 0.37 at 6, 12, and 18 months.

\subsubsection{AD Experiment 2}
\label{sec:case_study_AD_2}

In this second experiment, we analyze the performance of PROCOVA-MMRM vs MMRM when following the prospective implementation procedure described in Section III. To use the PROCOVA formula as a conservative estimate for PROCOVA-MMRM, we first needed to understand the Pearson correlation between the AD prognostic model predictions and actual outcomes in historical data that has similar characteristics to the AD Target Trial. Thus, a subset of the modeling data with baseline MMSE [14, 26] and age greater than 50 was used to estimate Pearson correlation coefficients of 0.267 and 0.361, for ADAS-Cog/11 and CDR-SB, respectively. This translates to an estimated sample size reduction of 7.1\% and 13.0\% using the PROCOVA formula with no deflation factor applied and assuming a gamma (inflation factor for the standard deviation) value of 1. This sample size reduction was then applied by randomly removing participants from both treatment and control arm of the AD Target Trial for PROCOVA-MMRM analysis. This process was repeated 1000 times and the average of the treatment effect variances with PROCOVA-MMRM was calculated. The estimated potential sample size reduction was accurate if the calculated treatment effect variance with PROCOVA-MMRM was equal to or less than the unadjusted MMRM estimate.

\begin{table}[H]
    \centering
    \begin{tabular} {|l|c|c|c|}
    \hline
         \textbf{Analysis Method} & \textbf{Sample Size} & \textbf{Treatment Effect Estimate} & \textbf{Estimated Variance} \\
          \hline
         Unadjusted MMRM & 402 & -0.035 & 1.024 \\ 
         PROCOVA-MMRM & 334 & 0.160 & 0.979 \\  
         \hline
    \end{tabular}
    \caption{Comparison of sample size, estimated treatment effect, and variance for ADAS-Cog11 at month 18 in reanalysis of the AD Target Trial.}
    \label{tab:AD-Experiment2-ADAS}
\end{table}

\begin{table}[H]
    \centering
    \begin{tabular} {|l|c|c|c|}
    \hline
         \textbf{Analysis Method} & \textbf{Sample Size} & \textbf{Treatment Effect Estimate} & \textbf{Estimated Variance} \\
          \hline
         Unadjusted MMRM & 402 & -0.092 & 0.113 \\ 
         PROCOVA-MMRM & 321 & -0.115 & 0.114 \\  
         \hline
    \end{tabular}
    \caption{Comparison of sample size, estimated treatment effect, and variance for CDR-SB at month 18 in reanalysis of the AD Target Trial.}
    \label{tab:AD-Experiment2-CDR}
\end{table}

Results in Table \ref{tab:AD-Experiment2-ADAS} shows that the treatment effect variances for PROCOVA-MMRM was lower than the unadjusted MMRM, indicating that the estimated sample size reduction was accurate. Similarly, Table \ref{tab:AD-Experiment2-CDR} shows very similar results between unadjusted MMRM and PROCOVA-MMRM using 13\% fewer subjects. We did not use a deflation factor for this exploratory analysis, but it can be useful in cases where the original estimate may be slightly optimistic. Regardless, this minor difference in variance would not have affected inference. Additionally, while the point estimates for the treatment effects were modified to some extent when prognostic score adjustment was applied, the changes were minimal relative to the size of the estimated standard errors. Overall, we demonstrate that the PROCOVA-MMRM model can be applied prospectively using an accurate estimate of the sample size reduction to the Target Trial. 

\subsection{Case Study 2 - Amyotrophic Lateral Sclerosis}
\label{sec:case_study_ALS}

Data from the ALS Target Trial \citep{Cudkowicz2014} were obtained from the National Institute of Neurological Disorders and Stroke (NINDS). The study consisted of Stage 1 (pharmacokinetics of ceftriaxone 2g and 4g over 7 days), Stage 2 (safety of the same doses over 20 weeks), and Stage 3 (efficacy and safety), where all patients on active treatment received 4g ceftriaxone regardless of their dose in the earlier stages. The final Stage 3 dataset comprised 513 participants; 340 in the ceftriaxone group and 173 in the placebo group. Participants were followed for 52 weeks after the randomization of the last participant.

The co-primary outcome measures used in the ALS Target Trial were the rate of functional decline as quantified by change in the ALS Functional Rating Scale-Revised (ALSFRS-R) over time, and the time to death, tracheostomy, or the start of permanent assisted ventilation, of which the former served as the primary outcome of interest. During Stages 1 and 2, a statistically significant difference was observed in the rate of functional decline (as measured via ALSFRS-R) for 4g ceftriaxone vs placebo. However, during Stage 3, no significant differences between the active and placebo groups were detected on either of the primary endpoints. For our re-analyses, we focused on the final phase 3 dataset of 513 participants; 340 in the ceftriaxone group and 173 in the placebo group (2:1 randomization ratio).

Historical de-identified ALS study data were obtained from four different sources and curated for model training. The Pooled Resource Open-Access ALS Clinical Trials (PRO-ACT) dataset contained data from 26 phase 2 and 3 RCTs encompassing 11675 trial participants. This data has been volunteered by members of the PRO-ACT Consortium, which was formed in 2011, Prize4Life, in collaboration with the Northeast ALS Consortium (NEALS), and with funding from the ALS Therapy Alliance. The NEALS dataset contained 5 clinical RCTs with 814 participants. The Pooled Resource Open-Access Clinical Research ALS (PRO-ACE) dataset included observational data for 4197 participants. The \cite{Witzel2021} dataset contained one observational and two interventional trials with 128 participants.

The data were processed to promote consistency and quality. Since the PRO-ACT dataset does not identify trials, it is possible that data from other sources is replicated in PRO-ACT. To correct for this issue, we applied a deduplication algorithm: participants were removed from the PRO-ACT dataset if a pre-defined subset of baseline and longitudinal variables (shared with other sources) matched or closely reproduced values observed in the NEALS dataset or in the ALS Target Trial dataset. In particular, we matched and removed all 513 subjects of the ALS Target Trial from PRO-ACT. Additional steps were taken to harmonize the data. Observations were set to standard units and data falling outside a predefined range were removed to promote quality and remove extreme outliers. For participants who dropped out from the study, the time to death was censored to the time of the last visit for which study data was reported. Participants were excluded if they had no available data for model training or if they died before study start (day 0). All 70 of the variables used to train the model were available for at least a subset of the clinical trials in the training dataset. Finally, longitudinal data was binned in time using a 28-days window centered around nominal visits with a 28 days cadence. The final curated dataset used for training and cross-validation consisted of 10,887 subjects.

\subsubsection{ALS Experiment 1}
\label{sec:case:study_ALS_1}

After fitting the prognostic models and obtaining prognostic scores for each study visit and participant in the ALS Target Trial, we analyzed the results for the treatment effect of ceftriaxone on the change in ALSFRS-R at 48 weeks using four approaches: MMRM; MMRM with covariate adjustment for BL ALSFRS-R only; PROCOVA-MMRM; and PROCOVA-MMRM with BL ALSFRS-R as an additional covariate. This experiment used the same number of participants and randomization ratio as the original study reported by \cite{Cudkowicz2014}. We compared the resulting point estimates and variance obtained with these four conditions for the effect of treatment on the changes in ALSFRS-R at 48 weeks (Table \ref{tab:ALS-Experiment1}). All data presented are heteroskedasticity consistent variance estimated with the Huber-White estimator.

\begin{table}[H]
    \centering
    \begin{tabular} {|l|c|c|}
    \hline
         \textbf{Analysis Method} & \textbf{Treatment Effect Estimate} & \textbf{Estimated Variance} \\
          \hline
         Unadjusted MMRM & 1.519 & 0.797 \\  
         MMRM + BL ALSFRS-R & 1.520 & 0.795 \\  
         PROCOVA-MMRM & 2.007 & 0.637 \\  
         PROCOVA-MMRM + BL ALSFRS-R & 2.069 & 0.627 \\   
         \hline
    \end{tabular}
    \caption{Estimated treatment effect and variance for ALSFRS-R at week 48 in reanalysis of the ALS Target Trial.}
    \label{tab:ALS-Experiment1}
\end{table}

We also evaluated how well the outcomes predicted by our prognostic model correlated with the subsequent study observations. The correlation between time-matched prognostic scores generated from baseline data and the study visit observations in the ALS Target Trial was 0.27, 0.37, 0.34, 0.38, 0.38 and 0.41 at weeks 8, 16, 24, 32, 40 and 48, respectively.

\subsubsection{ALS Experiment 2}
\label{sec:case_study_ALS_2}

In this second experiment, we analyze the performance of PROCOVA-MMRM vs MMRM when following the prospective implementation procedure as described in Section III. To use the PROCOVA formula as a conservative estimate for PROCOVA-MMRM, we first needed to understand the Pearson correlation between the ALS prognostic model predictions and actual outcomes in historical data that has similar characteristics to the ALS Target Trial. Thus, a subset of the modeling data with diagnosis within three years of baseline and 18 was used to obtain a Pearson correlation coefficient of 0.391 for ALSFRS-R. This translates to an estimated sample size reduction of 15.3\% using the PROCOVA formula with no deflation factor applied and assuming a gamma (inflation factor for the standard deviation) value of 1. This sample size reduction was then applied by randomly removing participants from both treatment and control arm of the ALS Target Trial for PROCOVA-MMRM analysis. This process was repeated 1000 times and the average of the treatment effect variances with PROCOVA-MMRM was calculated. The estimated potential sample size reduction was accurate if the calculated treatment effect variance with PROCOVA-MMRM was equal to or less than the unadjusted MMRM estimate.

\begin{table}[H]
    \centering
    \begin{tabular} {|l|c|c|c|}
    \hline
         \textbf{Analysis Method} & \textbf{Sample Size} & \textbf{Treatment Effect Estimate} & \textbf{Estimated Variance} \\
          \hline
         Unadjusted MMRM & 513 & 1.519 & 0.797 \\ 
         PROCOVA-MMRM & 427 & 1.997 & 0.753 \\  
         \hline
    \end{tabular}
    \caption{Comparison of sample size, estimated treatment effect, and variance for ALSFRS-R at week 48 in reanalysis of the ALS Target Trial.}
    \label{tab:ALS-Experiment2}
\end{table}

Results in Table \ref{tab:ALS-Experiment2} show that the endpoint treatment effect variance for PROCOVA-MMRM is lower than the unadjusted MMRM, indicating the same capacity for inference with a 15.3\% reduction in sample size. While the point estimates for the treatment effects were modified to some extent when prognostic score adjustment was applied, the changes were minimal relative to the size of the estimated standard errors. Similar to the AD case study, changes in treatment effect estimates were insignificant relative to the variance. This confirms that the PROCOVA-MMRM model can be applied prospectively using the PROCOVA formula to estimate an appropriate sample size reduction for the Target Trial.

\section{Simulation Studies of PROCOVA-MMRM}
\label{sec:simulation_studies}

We utilize several realistic simulated scenarios to demonstrate that PROCOVA-MMRM reduces variance and provides more precise estimates of treatment effects relative to MMRM (without adjustment for baseline covariates), while controlling type I error rate. Since the treatment effect is known, the difference between the true effect and the estimated result can be measured directly in various scenarios to characterize how treatment effect estimation differs between MMRM and PROCOVA-MMRM. Additional simulations and details on the simulation studies can be found in Appendix \ref{sec:appB}.

\subsection{Simulation Study Methods}
\label{sec:simulation_study_methods}

Assuming a treatment effect of -1.2, we simulated four different scenarios that model different situations encountered in clinical trials, and that enable us to probe the sensitivity of PROCOVA-MMRM to particular assumptions.

\begin{itemize}
    \item The Linear simulation describes a scenario in which the treatment effect is constant and the relationship between the time-matched prognostic scores and the expected outcomes is linear in both the treatment and control arms.
    \item The Additional Covariates simulation describes a scenario in which the relationships are the same as the Linear scenario, but three additional baseline covariates (ranging from 0.15-0.25 correlation with the outcome) have been added to the MMRM and PROCOVA-MMRM analyses.
    \item The Shifted simulation is similar to the Linear scenario, except that the time-matched prognostic scores have been systematically shifted. This simulates an event in which the prognostic scores may not fully apply to the Target Trial because of a population shift.
    \item The Heterogeneous simulation describes a scenario in which the conditional average effect $E[Y_{i,t}(1)-Y_{i,t}(0)|\mathbf{X}_i]=\mu_{1,t}(\mathbf{X}_i)-\mu_{0,t}(\mathbf{X}_i)$ is not constant (i.e., $E[Y_{i,t}(1)-Y_{i,t}(0)|\mathbf{X}_i]\neq{\mu_{1,t}-\mu_{0,t}}$). This represents an event where there is a nonconstant treatment effect.
\end{itemize}
The first two simulation scenarios fall under the assumptions in our theoretical results. Therefore, we expect PROCOVA-MMRM to perform well, so long as we use a prognostic model capable of capturing linear relationships. In contrast, the Heterogeneous scenario violates the constant additive treatment effect assumption of PROCOVA-MMRM, probing the sensitivity of PROCOVA-MMRM to such violations. Although the Shifted scenario does not violate any of the assumptions of PROCOVA-MMRM, a population shift in the trial participants’ characteristics may affect how strongly the time-matched prognostic scores are correlated with the expected outcomes in the Target Trial. Therefore, this scenario probes the sensitivity of PROCOVA-MMRM to the predictive performance of the trained prognostic model.

To compare treatment effect estimation precision, bias, confidence interval coverage and power between MMRM and PROCOVA-MMRM, we generated 2500 longitudinal datasets for each scenario with a total of 5 post-baseline measurements simulated for each participant. Observed outcomes were generated conditional on the prognostic score at a correlation of 0.5 with the final visit outcome. Missingness was assumed to follow the missing at random (MAR) assumption where the missingness is fully explained by baseline covariates only. Data were missing from approximately 30\% of participants, and the missingness was monotone. The alternative hypothesis was constructed such that there was 80\% power under the standard t-test. Additional details on the data generating process for each of the simulation scenarios are provided in Appendix \ref{sec:appB}.

\subsection{Simulation Study Results and Discussion}
\label{sec:simulation_study_results}

Table \ref{tab:simulations} presents the results obtained with MMRM and PROCOVA-MMRM in each of the scenarios described above, assuming a treatment effect of -1.2. Results include the point estimates of treatment effect with deviation from the true value, average variance of the estimate, the 95\% interval coverage probability, and the power.

\begin{table}[H]
    \centering
    \resizebox{\columnwidth}{!}{
\begin{tabular}{|l|c|c|c|c|c|}
\hline
 \textbf{Scenario} & \textbf{Treatment Effect Estimate} & \textbf{Bias} & \textbf{Average Variance} & \textbf{95\% CI Coverage} & \textbf{Power} \\
\hline
\multicolumn{4}{l}{\textbf{Linear}}\\
\hline
MMRM & -1.188 & 0.012 & 0.165 & 0.954 & 0.827 \\
\hline
PROCOVA-MMRM & -1.191 & 0.009 & 0.122 & 0.944 & 0.920 \\
\hline
\multicolumn{4}{l}{\textbf{Additional Covariates}}\\
\hline
MMRM & -1.191 & 0.009 & 0.141 & 0.945 & 0.886 \\
\hline
PROCOVA-MMRM & -1.195 & 0.005 & 0.098 & 0.946 & 0.967 \\
\hline
\multicolumn{4}{l}{\textbf{Shifted}}\\
\hline
MMRM & -1.209 & -0.009 & 0.165 & 0.947 & 0.841 \\
\hline
PROCOVA-MMRM & -1.211 & -0.011 & 0.145 & 0.954 & 0.887 \\
\hline
\multicolumn{4}{l}{\textbf{Heterogenous}}\\
\hline
MMRM & -1.173 & 0.027 & 0.166 & 0.951 & 0.814 \\
\hline
PROCOVA-MMRM & -1.176 & 0.024 & 0.123 & 0.949 & 0.910 \\
\hline
\end{tabular}
}
\caption{Simulation results with treatment effect = -1.2}
\label{tab:simulations}
\end{table}

In agreement with our case study results, PROCOVA-MMRM provides unbiased estimators of treatment effects and smaller average variance estimates that increase power while maintaining type I error control in all scenarios. The variances of the point estimators of the treatment effect under PROCOVA-MMRM is smaller than or equal to those obtained with MMRM. Our results show that PROCOVA-MMRM does not perform any worse than traditional approaches (nor does it lose its key statistical properties) even in circumstances when linear model assumptions (that are not specific to PROCOVA-MMRM) are challenged. Thus, PROCOVA-MMRM is a robust technique for estimating treatment effects from RCTs.

\section{Discussion}
\label{sec:discussion}

MMRM has been growing in popularity for longitudinal data analysis as it includes the intermediate observation data that is collected across study visits, simultaneously enhancing treatment estimation precision and handling the common RCT issue of missing data without explicit imputation \citep{Garcia2017,EMA_missingdata}. While covariate adjustment is a standard procedure for RCT analysis, PROCOVA enhances its benefit through incorporating AI-generated optimal adjustment covariates \citep{schuler_increasing_2021}. These predicted control outcomes, from individual baseline data, can be seamlessly integrated into a regulatory aligned framework for any clinical program where sufficient historical data is available to train and test a prognostic model \citep{EMAPROCOVA}. Here, we introduce PROCOVA-MMRM as a synergistic union of these two methods that is positioned to improve treatment effect estimation precision in longitudinal RCTs.

We provide mathematical, case study, and simulation evidence to support the advantageous properties of PROCOVA-MMRM. Namely, PROCOVA-MMRM yields unbiased estimators of treatment effects, smaller average variance estimates, increased power, and controlled type I error rates, highlighting its superiority over the unadjusted MMRM analysis. We also described an implementation of PROCOVA-MMRM that allows for prospective sample size reduction. This procedure allowed us to remove at random up to 15.3\% of clinical trial participants in a case study and still obtain the same results.

As described above, to enable the application of PROCOVA-MMRM, prognostic scores must be generated from the baseline characteristics of participants in the Target Trial using a prognostic model. Our procedure theoretically attains the largest decrease in variance when the prognostic model accurately predicts the expected control outcomes over time in the Target Trial population. However, some methods for fitting predictive models may overfit to the population in the training data, leading to a scenario in which the prognostic score has a much larger correlation with observed outcomes in the training dataset than in the Target Trial population. Therefore, to maximize the efficiencies gained through application of PROCOVA-MMRM, the prognostic model should be trained in a wide variety of data, but also perform well in a population similar to that of the Target Trial. While a lower performing model may result in a smaller efficiency gain, control of the type I error rate and unbiased treatment effect estimation will still be preserved with PROCOVA-MMRM in the large-sample setting.

Here we have outlined a streamlined implementation of PROCOVA-MMRM that utilizes the PROCOVA sample size formula as a conservative estimate for sample size reduction. Although this does not capture the maximum conversion from increased precision to prospective sample size reduction, it provides an additional buffer against underpowering the study and a logical path to validation of the prognostic model. Simulation studies may also provide insight into prospective estimation of sample size reduction/efficiency gain achievable through the use of any specific prognostic model with PROCOVA-MMRM, which may enable a greater reduction in sample size than the streamlined approach. Regardless of the method chosen for prospective implementation, the statistical properties of PROCOVA-MMRM are maintained.

We chose to assess PROCOVA-MMRM in ALS and AD here for several reasons, which included the availability of historical data, the unmet medical need, the widespread use of MMRM analysis in AD and ALS RCTs \citep{Sims2023,vanDyck2022,Haeberlein2022,Mintun2021,Ostrowitzki2017,Shefner2021}, and known challenges in recruitment and enrollment. However, the utility of PROCOVA-MMRM is not limited to incurable neurodegenerative diseases. Any clinical indication where a model with high predictive accuracy is available \citep{Walsh2024} can benefit from the integration of PROCOVA-MMRM into the design and analysis.

\section*{Financial Disclosure}
JLR, AS, RZ, and DB are equity-holding employees of Unlearn.AI, Inc., a company that creates digital twin generators to forecast patient outcomes.

\section*{Acknowledgements}
The authors would like to recognize Anna Vidovszky, Alyssa Vanderbeek, Stefanie Millar, Coco Kusiak, Ryan Douglas, and Eric Tramel for their ongoing support of this project, and David Walsh and David Miller for their previous work on this project.

Certain data used in the preparation of this submission were obtained from the University of California, San Diego Alzheimer’s Disease Cooperative Study (ADCS) Legacy database (National Institute on Aging Grant Number U19AG010483).

Certain data used in the preparation of this submission were obtained from the Alzheimer’s Disease Neuroimaging Initiative (ADNI) database. Data collection and sharing for this project was funded by the Alzheimer's Disease Neuroimaging Initiative (ADNI) (National Institutes of Health Grant U01 AG024904) and DOD ADNI (Department of Defense award number W81XWH-12-2-0012). ADNI is funded by the National Institute on Aging, the National Institute of Biomedical Imaging and Bioengineering, and through generous contributions from the following: AbbVie, Alzheimer’s Association; Alzheimer’s Drug Discovery Foundation; Araclon Biotech; BioClinica, Inc.; Biogen; Bristol-Myers Squibb Company; CereSpir, Inc.; Cogstate; Eisai Inc.; Elan Pharmaceuticals, Inc.; Eli Lilly and Company; EuroImmun; F. Hoffmann-La Roche Ltd and its affiliated company Genentech, Inc.; Fujirebio; GE Healthcare; IXICO Ltd.; Janssen Alzheimer Immunotherapy Research \& Development, LLC.; Johnson \& Johnson Pharmaceutical Research \& Development LLC.; Lumosity; Lundbeck; Merck \& Co., Inc.; Meso Scale Diagnostics, LLC.; NeuroRx Research; Neurotrack Technologies; Novartis Pharmaceuticals Corporation; Pfizer Inc.; Piramal Imaging; Servier; Takeda Pharmaceutical Company; and Transition Therapeutics. The Canadian Institutes of Health Research is providing funds to support ADNI clinical sites in Canada. Private sector contributions are facilitated by the Foundation for the National Institutes of Health (www.fnih.org). The grantee organization is the Northern California Institute for Research and Education, and the study is coordinated by the Alzheimer’s Therapeutic Research Institute at the University of Southern California. ADNI data are disseminated by the Laboratory for NeuroImaging at the University of Southern California.

Certain data used in the preparation of this submission were obtained from the Critical Path for Alzheimer's Disease (CPAD) database. In 2008, Critical Path Institute, in collaboration with the Engelberg Center for Health Care Reform at the Brookings Institution, formed the Coalition Against Major Diseases (CAMD), which was then renamed to CPAD in 2018. The Coalition brings together patient groups, biopharmaceutical companies, and scientists from academia, the U.S. Food and Drug Administration (FDA), the European Medicines Agency (EMA), the National Institute of Neurological Disorders and Stroke (NINDS), and the National Institute on Aging (NIA). CPAD currently includes over 200 scientists, drug development and regulatory agency professionals, from member and non-member organizations. The data available in the CPAD database has been volunteered by CPAD member companies and non-member organizations.

Certain data used in this submission were provided by the Knight Alzheimer Disease Research Center and OASIS-4: Clinical Cohort (Principal Investigators: T. Benzinger, L. Koenig, P. LaMontagne).

Certain data used in preparation of this article were obtained from the EPAD LCS data set V.IMI, doi:10.34688/epadlcs\_v.imi\_20.10.30. This work used data from the EPAD project which received support from the EU/EFPIA Innovative Medicines Initiative Joint Undertaking EPAD grant agreement n° 115736 and an Alzheimer’s Association Grant (SG-21-818099-EPAD). EPAD LCS is registered at www.clinicaltrials.gov Identifier: NCT02804789.

Certain data used in this submission were obtained from the NACC database. The NACC database is funded by NIA/NIH Grant U24 AG072122. NACC data are contributed by the NIA-funded ADCs: P50 AG005131 (PI James Brewer, MD, PhD); P50 AG005133 (PI Oscar Lopez, MD); P50 AG005134 (PI Bradley Hyman, MD, PhD); P50 AG005136 (PI Thomas Grabowski, MD); P50 AG005138 (PI Mary Sano, PhD); P50 AG005142 (PI Helena Chui, MD); P50 AG005146 (PI Marilyn Albert, PhD); P50 AG005681 (PI John Morris, MD); P30 AG008017 (PI Jeffrey Kaye, MD); P30 AG008051 (PI Thomas Wisniewski, MD); P50 AG008702 (PI Scott Small, MD); P30 AG010124 (PI John Trojanowski, MD, PhD); P30 AG010129 (PI Charles DeCarli, MD); P30 AG010133 (PI Andrew Saykin, PsyD); P30 AG010161 (PI David Bennett, MD); P30 AG012300 (PI Roger Rosenberg, MD); P30 AG013846 (PI Neil Kowall, MD); P30 AG013854 (PI Robert Vassar, PhD); P50 AG016573 (PI Frank LaFerla, PhD); P50 AG016574 (PI Ronald Petersen, MD, PhD); P30 AG019610 (PI Eric Reiman, MD); P50 AG023501 (PI Bruce Miller, MD); P50 AG025688 (PI Allan Levey, MD, PhD); P30 AG028383 (PI Linda Van Eldik, PhD); P50 AG033514 (PI Sanjay Asthana, MD, FRCP); P30 AG035982 (PI Russell Swerdlow, MD); P50 AG047266 (PI Todd Golde, MD, PhD); P50 AG047270 (PI Stephen Strittmatter, MD, PhD); P50 AG047366 (PI Victor Henderson, MD, MS); P30 AG049638 (PI Suzanne Craft, PhD); P30 AG053760 (PI Henry Paulson, MD, PhD); P30 AG066546 (PI Sudha Seshadri, MD); P20 AG068024 (PI Erik Roberson, MD, PhD); P20 AG068053 (PI Marwan Sabbagh, MD); P20 AG068077 (PI Gary Rosenberg, MD); P20 AG068082 (PI Angela Jefferson, PhD); P30 AG072958 (PI Heather Whitson, MD); and P30 AG072959 (PI James Leverenz, MD).

Certain data used in this submission were obtained from the National Institute of Neurological Diseases and Stroke’s Archived Clinical Research data (Title: Clinical Trial Ceftriaxone in Subjects with Amyotrophic Lateral Sclerosis, PI: E. Cudkowicz, MD, Funding: NINDS 5 U01-NS-049640) received from the Archived Clinical Research Dataset web site (ninds.nih.gov/currentresearch/research-funded-ninds/clinical-research/archived-clinical-research-datasets).

Certain data used in this submission were obtained from the PRO-ACT Database, where they have been volunteered by members of the PRO-ACT Consortium: ALS Therapy Alliance, Cytokinetics, Inc., Amylyx Pharmaceuticals, Inc., Knopp Biosciences, Neuraltus Pharmaceuticals, Inc., Neurological Clinical Research Institute, MGH, Northeast ALS Consortium, Novartis, Prize4Life Israel, Regeneron Pharmaceuticals, Inc., Sanofi, Teva Pharmaceutical Industries, Ltd., and The ALS Association.

A subset of the data used in preparation of this submission were obtained from NEALS. We acknowledge the NEALS Biorepository for providing all or part of the biofluids from the ALS, healthy controls, and non-ALS neurological controls used in this study.

A subset of the data used in preparation of this document were obtained from the authors of 'Neurofilament light and heterogeneity of disease progression in amyotrophic lateral sclerosis'. We acknowledge the authors Simon Witzel, Felix Frauhammer, and Petra Steinacker for their work (Witzel et al 2021). The data used in this project is attributed under the Creative Commons license: https://creativecommons.org/licenses/by/4.0/

Certain data used in the preparation of this submission were obtained from the PRO-ACE Database. The data available in the PRO-ACE Database have been volunteered by PRO-ACE Consortium members.

\bibliographystyle{apalike}
\bibliography{PROCOVA-MMRM}

\renewcommand{\thesubsection}{\Alph{subsection}}

\pagebreak

\section{Appendix}
\label{sec:appendix}

\subsection{Missing Data Assumptions}
\label{sec:missingness_assumptions}

In order to compare the precisions of different methods for repeated measures analyses in the case of missing data, we generally assume that the data are missing completely at random (MCAR) so as to have a fair basis for comparison. For example, if the data were missing at random (MAR), then the traditional MMRM analysis that does not involve any adjustment would correspond to an analysis under a missing not at random (MNAR) mechanism, and would likely yield biased treatment effect inferences. As such, it would be difficult to justify a fair comparison of PROCOVA-MMRM (or any MMRM method that adjusts for baseline covariates) to the traditional MMRM analysis, because the latter would be biased whereas the former could be unbiased. We shall also assume that the missing data pattern is monotone.

\subsection{Relating Sample Size Reduction to the Effective Sample Size}
\label{sec:appA}

We provide a simple proof, based on the concept of the effective sample size (ESS), to demonstrate how the sample size for a repeated measures trial calculated according to the PROCOVA formula would be a conservative estimate of the sample size that would be necessary for a PROCOVA-MMRM analysis.

The ESS is a metric that can capture the sample size efficiency of one method of estimation versus a benchmark method. An explanation of ESS is provided in the \citet[p.~39--40]{food_and_drug_administration_bayesian_2010} \emph{Guidance for the Use of Bayesian Statistics in Medical Device Clinical Trials}. It can be interpreted for a Bayesian method as a measure of the number of patients that are ``borrowed'' from a historical trial in the specification of the prior distribution for the Bayesian analysis. To formally define this quantity, let $V_{\mathrm{benchmark}}$ denote the variance of the treatment effect estimator under a benchmark method, and $V_{\mathrm{new}}$ denote the variance of the treatment effect estimator under a new method. Then the ESS is defined as

\begin{equation}
\label{eq:ESS_general_definition}
\mathrm{ESS} = \frac{NV_{\mathrm{benchmark}}}{V_{\mathrm{new}}},
\end{equation}

\noindent where $N$ denotes the trial sample size. The core of this definition is the ratio of the variances, which is our focus as well.

The ESS can also be related to the sample size reduction of a new method compared to a benchmark method. This broader scope of the ESS is demonstrated by means of a Taylor series expansion of the sample size formula for the new method. Let $N_0$ denote the number of control participants and $N_1$ denote the number of treated participants in a trial. We consider a sample size formula for a new method that is just a function of $N_0$, i.e., the number of treated participants $N_1$ is held fixed throughout. Also, let $f: \mathbb{R}^+ \rightarrow \mathbb{R}^+$ denote the function whose input is the control group sample size $N_0$, and whose output is the \emph{precision} of the treatment effect estimator under the new method. We consider precision (the inverse of variance) because of the comparative ease in considering the derivatives $f', f''$, etc., in this case.

Suppose $n_0$ is the control arm sample size that was identified for the trial under the analysis by the benchmark method. For example, when the benchmark analysis is ANOVA, $n_0$ would denote the control arm sample size under which ANOVA would have power $80\%$. We assume that an expression exists for the variance of the treatment effect estimator under the benchmark method at the sample size $n_0$. Let $V_{\mathrm{benchmark}}$ denote the variance of the treatment effect estimator under the benchmark method with a control arm sample size of $n_0$.

The Taylor series expansion for $f$ at $n_0$ is

\begin{equation}
\label{eq:taylor_series}
f(N_0) = f(n_0) + (N_0-n_0)f'(n_0) + \frac{1}{2}(N_0-n_0)^2f''(n_0) + \cdots.
\end{equation}

\noindent Suppose $f''(n_0), f^{(3)}(n_0)$, and the other higher-order derivatives in this expansion are negligible. Then 

\begin{equation}
\label{eq:taylor_series_approximation}
f(N_0) \approx f(n_0) + (N_0-n_0)f'(n_0).
\end{equation}

\noindent We use equation (\ref{eq:taylor_series_approximation}) to find the approximate value of $N_0$ such that $f(N_0) = V_{\mathrm{benchmark}}^{-1}$:

\begin{equation}
\label{eq:taylor_series_root}
N_0 \approx n_0 - \left \{ \frac{V_{\mathrm{benchmark}}f(n_0) - 1}{V_{\mathrm{benchmark}}f'(n_0)} \right \}.    
\end{equation}

\noindent Let $f(n_0) = V_{\mathrm{new}}^{-1}$ denote the precision of the new method under the control arm sample size of $n_0$. Then

\begin{align}
\label{eq:taylor_series_solution}
N_0 &\approx n_0 - \left \{ \frac{V_{\mathrm{benchmark}}/V_{\mathrm{new}} - 1}{V_{\mathrm{benchmark}}f'(n_0)} \right \} \\
&\approx n_0 - \left \{ \frac{(\mathrm{ESS} - N)}{NV_{\mathrm{benchmark}}f'(n_0)} \right \},    
\end{align}

where $N = n_0 + N_1$. Thus, $N_0$ can be approximated using the $\mathrm{ESS}$. It is important to note the following points that are involved in this derivation.

\begin{itemize}

\item We are considering the case in which $V_{\mathrm{benchmark}}/V_{\mathrm{new}} > 1$ under the control arm sample size $n_0$ and treatment arm sample size $N_1$.

\item We assume $f'(n_0) > 0$ for all $n_0$ and $f'(n_0) \rightarrow 0$ as $n_0 \rightarrow \infty$.

      \begin{itemize}

      \item For a well-behaved treatment effect estimator, we can expect its precision to increase monotonically as a function of the sample size.

      \item Furthermore, we can expect the rate of increase for the precision to diminish, and ultimately converge to zero, as the sample size increases.
      
      \end{itemize}

\item As ESS increases, $N_0$ decreases (i.e., there is greater sample size reduction).
    
\end{itemize}

\subsubsection{Simple Proof for the Effective Sample Size for PROCOVA-MMRM Versus PROCOVA}
\label{sec:ESS_procova_mmrm}

Let $\boldsymbol{\beta} = \begin{pmatrix} \boldsymbol{\beta}_0 \\ \boldsymbol{\beta}_w \\ \boldsymbol{\beta}_x \end{pmatrix}$ denote the vector of regression coefficients for PROCOVA-MMRM, and $\boldsymbol{\phi}$ be the vector containing all the parameters in the participants' covariance matrices $\boldsymbol{\Psi}$. Both the maximum likelihood estimator (MLE) and restricted maximum likelihood estimator (REML) for $\begin{pmatrix} \boldsymbol{\beta} \\ \boldsymbol{\phi} \end{pmatrix}$ are obtained by first identifying the vector $\hat{\boldsymbol{\phi}}$  that maximizes the respective profile likelihood, and then identifying $\boldsymbol{\beta}$ via
\begin{equation}
\label{eq:beta_estimates_REML}
\hat{\boldsymbol{\beta}} \left ( \hat{\boldsymbol{\phi}} \right ) = \left ( \mathbf{X}^{\mathsf{T}} \hat{\boldsymbol{\Omega}}^{-1} \mathbf{X} \right )^{-1} \mathbf{X}^{\mathsf{T}} \hat{\boldsymbol{\Omega}}^{-1} \mathbf{y},
\end{equation}
where $\mathbf{X}$ is the row-wise concatenation of the predictor matrices for the participants, and $\boldsymbol{\Omega}$ is the $\left ( \displaystyle \sum_{i=1}^N T_i \right ) \times \left ( \displaystyle \sum_{i=1}^N T_i \right )$ block-diagonal matrix with diagonal components $\mathbf{R}_1, \ldots, \mathbf{R}_N$. For the purposes of simplifying the proof, for now we consider the case that $\boldsymbol{\Omega}$ is known and focus on $\hat{\boldsymbol{\beta}} \left ( \boldsymbol{\phi} \right )$. The covariance matrix of this estimator is
\begin{equation}
\label{eq:procova_mmrm_covariance_matrix}
\left ( \mathbf{X}^{\mathsf{T}} \boldsymbol{\Omega}^{-1} \mathbf{X} \right )^{-1} = \left \{ \sum_{i=1}^N \left ( \mathbf{X}_i^{\mathsf{T}} \mathbf{R}_i^{-1} \mathbf{X}_i \right ) \right \}^{-1},    
\end{equation}
where $\mathbf{X}_i$ is the $T_i \times 3T$ predictor matrix for participant $i$. The precision matrix is thus
\begin{equation}
\label{eq:procova_mmrm_precision_matrix}
\sum_{i=1}^N \left ( \mathbf{X}_i^{\mathsf{T}} \mathbf{R}_i^{-1} \mathbf{X}_i \right ).    
\end{equation}
We note that
\[
\mathbf{X}_i^{\mathsf{T}} \mathbf{R}_i^{-1} \mathbf{X}_i = \begin{pmatrix} \mathbf{R}_i^{-1} & \mathbf{0} & w_i \mathbf{R}_i^{-1} & \mathbf{0} & \mathbf{R}_i^{-1} \mathbf{D}_i & \mathbf{0} \\ \mathbf{0} & \mathbf{0} & \mathbf{0} & \mathbf{0} & \mathbf{0} & \mathbf{0} \\ w_i \mathbf{R}_i^{-1} & \mathbf{0} & w_i \mathbf{R}_i^{-1} & \mathbf{0} & w_i \mathbf{R}_i^{-1} \mathbf{D}_i & \mathbf{0} \\ \mathbf{0} & \mathbf{0} & \mathbf{0} & \mathbf{0} & \mathbf{0} & \mathbf{0} \\ \mathbf{D}_i^{\mathsf{T}} \mathbf{R}_i^{-1} & \mathbf{0} & w_i \mathbf{D}_i^{\mathsf{T}} \mathbf{R}_i^{-1} & \mathbf{0} & \mathbf{D}_i^{\mathsf{T}} \mathbf{R}_i^{-1} \mathbf{D}_i & \mathbf{0} \\ \mathbf{0} & \mathbf{0} & \mathbf{0} & \mathbf{0} & \mathbf{0} & \mathbf{0} \end{pmatrix}
\]
where $\mathbf{D}_i = \mathrm{diag} \left ( x_{i,1}, \ldots, x_{i,T_i} \right )$ is the $T_i \times T_i$ diagonal matrix consisting of the $T_i$ prognostic scores for participant $i$, and the $\mathbf{0}$ matrices correspond to the time points that are missing for participant $i$. 

Let $\mathcal{S}$ denote those participants who have complete outcomes. Then the component of the precision matrix above corresponding to these subjects is
\[
\sum_{i \in \mathcal{S}} \left ( \mathbf{X}_i^{\mathsf{T}} \mathbf{R}_i^{-1} \mathbf{X}_i \right ) = \begin{pmatrix} N\boldsymbol{\Psi}^{-1} & N_1 \boldsymbol{\Psi}^{-1} & \boldsymbol{\Psi}^{-1} \left ( \displaystyle \sum_{i \in \mathcal{S}} \mathbf{D}_i \right ) \\ N_1 \boldsymbol{\Psi}^{-1} & N_1 \boldsymbol{\Psi}^{-1} &  \boldsymbol{\Psi}^{-1} \left ( \displaystyle \sum_{i \in \mathcal{S}} w_i \mathbf{D}_i \right ) \\ \left ( \displaystyle \sum_{i \in \mathcal{S}} \mathbf{D}_i^{\mathsf{T}} \right ) \boldsymbol{\Psi}^{-1} & \left ( \displaystyle \sum_{i \in \mathcal{S}} w_i \mathbf{D}_i^{\mathsf{T}} \right ) \boldsymbol{\Psi}^{-1} & \left ( \displaystyle \sum_{i \in \mathcal{S}} \mathbf{D}_i^{\mathsf{T}} \boldsymbol{\Psi}^{-1} \mathbf{D}_i \right ) \end{pmatrix}
\]

Finally, as $\sum_{i=1}^N \left ( \mathbf{X}_i^{\mathsf{T}} \mathbf{R}_i^{-1} \mathbf{X}_i \right ) \geq \sum_{i \in \mathcal{S}} \left ( \mathbf{X}_i^{\mathsf{T}} \mathbf{R}_i^{-1} \mathbf{X}_i \right )$ in terms of positive semi-definiteness, we then have that $\left \{ \sum_{i=1}^N \left ( \mathbf{X}_i^{\mathsf{T}} \mathbf{R}_i^{-1} \mathbf{X}_i \right ) \right \}^{-1} \leq \left \{ \sum_{i \in \mathcal{S}} \left ( \mathbf{X}_i^{\mathsf{T}} \mathbf{R}_i^{-1} \mathbf{X}_i \right ) \right \}^{-1}$ in terms of positive semidefiniteness. The latter inequality enables us to establish the inequality between the variance of the treatment effect estimator under PROCOVA-MMRM to the treatment effect estimator under PROCOVA.

\subsection{Details of Simulation Studies}
\label{sec:appB}

In our simulation studies, we considered a RCT where the endpoint is continuous and each patient has measurements of the endpoint at 5 total endpoints. The data were analyzed using a MMRM model using all available measurements. Data generation was simulated using the block matrices of a 10x10 covariance matrix where the upper left (Quadrant II) block matrix was the 5x5 covariance matrix of observed outcomes, the bottom right (Quadrant IV) block matrix was the 5x5 covariance matrix of the time-matched prognostic scores, and the off diagonal (Quadrants I, III) block matrices was the 5x5 covariance matrix of the observed outcome and time-matched prognostic scores. 

The time-matched prognostic scores were first generated from a multivariate normal distribution with mean 0 and covariance defined by the 5x5 block matrix in Quadrant IV. The observed outcomes were generated conditional on the time-matched prognostic scores using a multivariate normal distribution with conditional mean and covariance derived from the block covariance matrices and mean prognostic score. In all simulation scenarios except the shifted scenario, the mean deviation of the time-matched prognostic scores from the observed control outcome was 0. Missingness was simulated as a function of 3 baseline covariates, resulting in approximately 30\% missingness at the final timepoint. Missingness was also assumed to be monotone. A total of 1,000 participants before dropout was simulated in each arm, and each scenario was repeated 2,500 times.

For the analysis, the prognostic score and time interaction was included to account for the time-matched prognostic scores. An unstructured covariance matrix was used to model the within-participant correlations. The estimated Least Square Means (LSM) and variance estimator of the LSM were used to assess the statistical properties. The Kenward-Roger approximation was used to adjust the degrees of freedom for hypothesis testing. The longitudinal datasets and analysis were completed in R using the “mmrm” package. 

Further details regarding the scenarios:
\begin{itemize}
    \item Linear: Follows exactly the text above.
    \item Shifted: A random shift was added to the time-matched prognostic scores, mimicking the scenario where there are population differences between the historical training set and Target Trial. The time-matched prognostic scores at each timepoint were shifted by a normal distribution with mean vector -.3, -.5, -1, -2, -2.5 and standard deviation 3, 4 ,6, 7, 8. Adding this shift changes the covariance matrix of the prognostic score and observed outcomes. While the original simulation targeted a final correlation of 0.5, the resulting mean correlation was only 0.37.
    \item Additional Covariates: Three continuous baseline covariates were generated with moderate prognostic ability. Each baseline covariate had an approximate correlation of 0.25 with the outcome. Therefore, the covariance matrix to generate all observed in this scenario was a 13x13 matrix. Baseline covariate and time interactions were included in these analyses.
    \item Heterogeneous: A non-constant treatment effect was simulated by generating outcomes that were conditional on the baseline covariate by treatment interaction. Three baseline covariates were used, and the true treatment effect was estimated by generating the complete set of potential outcomes under the hypothetical scenario for all participants. The baseline by treatment interaction terms were not included in the analysis model.
\end{itemize}

Here we provide additional simulation results for the aforementioned scenarios in the condition that the null hypothesis is correct.

\begin{table}[H]
    \centering
    \resizebox{\columnwidth}{!}{
\begin{tabular}{|l|c|c|c|c|c|}
\hline
 \textbf{Scenario} & \textbf{Treatment Effect Estimate} & \textbf{Bias} & \textbf{Average Variance} & \textbf{95\% CI Coverage} & \textbf{Type I Error} \\
\hline
\multicolumn{4}{l}{\textbf{Linear}}\\
\hline
MMRM & 0.014 & 0.014 & 0.165 & 0.954 & 0.046 \\
\hline
PROCOVA-MMRM & 0.009 & 0.009 & 0.122 & 0.944 & 0.054 \\
\hline
\multicolumn{4}{l}{\textbf{Additional Covariates}}\\
\hline
MMRM & 0.009 & 0.009 & 0.141 & 0.945 & 0.055 \\
\hline
PROCOVA-MMRM & 0.005 & 0.005 & 0.098 & 0.946 & 0.054 \\
\hline
\multicolumn{4}{l}{\textbf{Shifted}}\\
\hline
MMRM & -0.009 & -0.009 & 0.165 & 0.947 & 0.053 \\
\hline
PROCOVA-MMRM & -0.011 & -0.011 & 0.146 & 0.954 & 0.046 \\
\hline
\multicolumn{4}{l}{\textbf{Heterogenous}}\\
\hline
MMRM & 0.007 & 0.007 & 0.165 & 0.953 & 0.048 \\
\hline
PROCOVA-MMRM & 0.005 & 0.005 & 0.122 & 0.946 & 0.053 \\
\hline
\end{tabular}
}
    \caption{Simulation results with treatment effect = 0}
    \label{tab:simulations0}
\end{table}

In this simulation setting, the correlation between the time-matched prognostic scores and observed outcomes was easily controlled using the block covariance matrices. For all prior simulations (Tables \ref{tab:simulations} and \ref{tab:simulations0}), we assumed a correlation of 0.5 between the prognostic model predictions and observed outcomes. Here we provide additional linear simulation results using a correlation value of 0.1 to demonstrate how a lower performing prognostic model affects treatment effect estimation in PROCOVA-MMRM.

\begin{table}[H]
    \centering
    \resizebox{\columnwidth}{!}{
\begin{tabular}{|l|c|c|c|c|c|}
\hline
 \textbf{Condition} & \textbf{Treatment Effect Estimate} & \textbf{Bias} & \textbf{Average Variance} & \textbf{95\% CI Coverage} & \textbf{Null Rejection Probability} \\
\hline
\multicolumn{2}{l}{\textbf{H1 Linear}}\\
\hline
MMRM & -1.191 & 0.009 & 0.164 & 0.944 & 0.833 \\
\hline
PROCOVA-MMRM & -1.191 & 0.009 & 0.163 & 0.944 & 0.836 \\
\hline
\multicolumn{2}{l}{\textbf{H0 Linear}}\\
\hline
MMRM & 0.009 & 0.009 & 0.164 & 0.944 & 0.056 \\
\hline
PROCOVA-MMRM & 0.009 & 0.009 & 0.163 & 0.944 & 0.054 \\
\hline
\hline
\end{tabular}
}
    \caption{Additional Linear Simulation Results for a Less Correlated Prognostic Model}
    \label{tab:simulationsapp}
\end{table}

The H1 linear condition in Table \ref{tab:simulationsapp} corresponds with the linear scenario results in Table \ref{tab:simulations} of the main text (i.e., treatment effect), and the H0 Linear condition corresponds with the linear scenario results in Table \ref{tab:simulations0} of the appendix (i.e., no treatment effect). Note that the unadjusted MMRM results have also changed due to the change in data generating mechanism for the expected outcome. Regardless if a correlation of 0.1 or 0.5 was used, PROCOVA-MMRM returned similar results for point estimate of treatment effect, 95\% confidence interval coverage, and type I error rate. As expected, the lower performing model showed much less variance reduction than the higher performing model.

\end{document}